\documentstyle[12pt,aasms4]{article}
\def\lapprox{\hbox{\lower .8ex\hbox{$\,\buildrel < \over\sim\,$}}}
\def\gapprox{\hbox{\lower .8ex\hbox{$\,\buildrel > \over\sim\,$}}}

\begin{document}

\title{The binary progenitor of Tycho Brahe's 1572 supernova}

\bigskip

\noindent
Pilar Ruiz-Lapuente$^{1,2}$, Fernando Comeron$^{3}$, Javier M\'endez$^{1,4}$,
Ramon Canal$^{1}$,  Stephen J. Smartt$^{5}$, Alexei V. Filippenko$^{6}$,
Robert L. Kurucz$^{7}$, Ryan Chornock$^{6}$,  
Ryan J. Foley$^{6}$, Vallery Stanishev$^{8}$, \& Rodrigo Ibata$^{9}$

\noindent
$^{1}$ Department of Astronomy, University of Barcelona, Marti i Franques 1, 
E--08028 Barcelona, Spain

\noindent
$^{2}$ Max-Planck Institut f\"ur Astrophysik, Karl-Schwarzschild-Strasse 1,
 85748 Garching, Germany

\noindent
$^{3}$ European Southern Observatory, Karl-Schwarzschild-Strasse 2, 
85748 Garching, Germany


\noindent
$^{4}$ Isaac Newton Group, P.O. Box 321, Santa Cruz de La Palma, 
Tenerife, Canary Islands,
E--38780 Spain 

\noindent
$^{5}$ Institute of Astronomy, University of Cambridge, Madingley Road,
Cambridge CB3 0HA, UK$^{*}$

\noindent
$^{6}$ Department of Astronomy, 601 Campbell Hall, University of
 California, Berkeley, CA 94720--3411 USA 

\noindent
$^{7}$ Harvard-Smithsonian Center for Astrophysics, Cambridge, 
MA 02138, USA

\noindent
$^{8}$ Department of Physics, Stockholm University, AlbaNova
University Center, SE-108 91 Stockholm, Sweden 

\noindent
$^{9}$ Observatoire de Strasbourg, 11, rue de l'Universit\'e, F-67000
 Strasbourg, France

\noindent
$^{*}$Present address: Department of Physics, Queen's University Belfast,
Belfast BT7 1NN, Northern Ireland, UK

\bigskip

{\bf 
The brightness of type Ia supernovae, and their homogeneity as a class, 
makes them powerful tools in cosmology, yet little is known about the 
progenitor systems of these explosions. They are thought to arise when a 
white dwarf accretes matter from a companion star, is compressed and undergoes
a thermonuclear explosion$^{1-3}$. Unless the companion star is another white
dwarf (in which case it should be destroyed by the mass-transfer process 
itself), it should survive and show distinguishing properties. Tycho's 
supernova$^{4-5}$ is one of the only two type Ia supernovae observed in our
Galaxy, and so provides an opportunity to address observationally the 
identification of the surviving companion. Here we report a survey of the 
central region of its remnant, around the position of the explosion, which 
excludes red giants as the mass donor of the exploding white dwarf. We found
a type G0--G2 star, similar to our Sun in surface temperature and luminosity 
(but lower surface gravity), moving at more than three times the mean 
velocity of the stars at that distance, which appears to be the surviving 
companion of the supernova.}

Tycho Brahe's supernova (that is, SN 1572) is one of the only two supernovae 
observed in our Galaxy that are thought to have been of type Ia (the other 
having been SN 1006) as revealed by the light curve, radio emission 
and X--ray spectra$^{4-7}$.
 
The field that contained Tycho's supernova, relatively devoid of background
stars, is favourable for searching for any surviving companion. With a 
Galactic latitude $b = +1.4^\circ$, Tycho's supernova lies 59--78 pc above 
the Galactic plane. The stars in that direction show a consistent pattern of 
radial velocities with a mean value of $-30$ km s$^{-1}$ at 3 kpc. The 
predictions of how the companion star would look after the impact, if there 
is any companion, depend on what the star actually is. The star could be in 
any evolutionary stage before the explosion: main sequence, subgiant or red 
giant$^{1-3}$. The most salient feature of the surviving companion star 
should be peculiar velocities with respect to the average motion of the other 
stars at the same location in the Galaxy (mainly due to disruption of the 
binary)$^{8}$, detectable through radial--velocity measurements, and perhaps 
also signs of the impact of the supernova ejecta. The latter can be twofold. 
First, mass should have been stripped from the companion and thermal energy 
injected into it, possibly leading to expansion of the stellar envelope that 
would make the star have a lower surface gravity. Second, depending on the 
interaction with the ejected material, the surface of the star could be 
contaminated by the slowest--moving ejecta (made of Fe and Ni isotopes). If 
the companion's stellar envelope is radiative, such a contamination could be
detectable through abundance measurements. Therefore, the observations have 
been designed along these lines. The star most likely to have been the mass 
donor of SN 1572 has to show a multiple coincidence: being at the distance of 
SN 1572, it has to show an unusual radial velocity in comparison to the stars 
at the same location (much above the velocity dispersion for its spectral 
type), and have stellar parameters consistent with being struck by the SN 
explosion.  It should also lie near the remnant centre (that is, within our 
search radius). 

The distance to SN 1572 inferred from the expansion of the radio shell and by 
other methods lies around 3 kpc ($2.83 \pm 0.79$ kpc)$^{9}$. Such a distance, 
and the light--curve shape of SN 1572, are consistent with it being a normal 
type Ia supernova in luminosity, as those commonly found in cosmological 
searches$^{9}$. Given the age of the supernova remnant (SNR; just 432 yr) 
and the lower limit to its distance, any possible companion, even if it moved 
at a speed of 300 km s$^{-1}$, could not be farther than 0.15 arcmin (9.1 
arcsec) from its position at the time of the explosion$^{8,10}$. But the 
search radius significantly expands owing to the uncertainty in the derived 
centre of the SNR (see Fig. 1).

We have analysed the stars within a circle of 0.65 arcmin radius, centred 
on the Chandra X-ray Observatory coordinates for the centre of the SNR, 
up to an apparent visual magnitude $V = 22$ (Figure 1, Figure 2, Table 1, 
and Supplementary Tables 1--3). 

All but one of the stars found are either main-sequence stars (luminosity 
class V) with spectral types A4--K3 or giant stars (luminosity class III) 
with spectral types G0--K3. 

\noindent
Red--giant stars are possible companions of type Ia supernovae. Masses in 
the range 0.9--1.5 solar masses (0.9--1.5 M$_{\odot}$) would be the most 
favourable cases$^{11}$. Red giants are well represented in the sample, but 
none of them passes the tests for being a viable candidate. They are at 
distances incompatible with that of the supernova. The only giant relatively 
close to the distance of SN 1572 is Tycho A (Fig. 3), but it is closer than 
SN 1572 and shows no peculiarities in velocity, spectral type, or metallicity. 
Main-sequence stars are also viable companions of type Ia supernovae. Close 
binaries with 2 to  3.5 M$_{\odot}$  main--sequence or subgiant companions 
have indeed been suggested as one class of systems able to produce type Ia 
supernovae$^{12}$. Among systems containing a main--sequence star, recurrent 
novae have been pointed out as possible progenitors$^{13}$. Stripping of mass 
from the impact of the ejecta on this type of companion is also 
expected$^{8,14}$. Another consequence of the impact should be to puff up the 
star and dramatically increase its luminosity. The size and luminosity would 
later return to their equilibrium values for a star with the new decreased 
mass$^{8,14,15}$. Peculiar velocities should be highest (200--300 km 
s$^{-1}$) in the case of main--sequence companions (orbital separations at 
the time of explosion are shortest), but the measured values of radial 
velocity ($v_r$) for the main--sequence stars observed are not particularly 
high. The surface abundances are compatible with solar values. Other 
main--sequence stars (see Fig. 1, Table 1 and Supplementary Table 3) are 
found at wider separations from the geometrical centre, but they have $v_{r}$
values within the range corresponding to their respective distances (see 
Supplementary Discussion). 

\bigskip
  
We have found a subgiant star (`Tycho G') with lower surface gravity than 
that of main--sequence stars but higher surface gravity than that of red 
giants, which moves fast in comparison to the mean radial velocities of stars 
around it, and fits well the expectations for distance, reddening and 
velocity. Comparison of the Tycho G spectrum covering a wide wavelength range 
(3,180--9,400 \AA) with templates$^{16}$, after dereddening by $E(B-V) 
\approx 0.6$ mag, gives a best fit for an effective temperature $T_{eff}$ = 
5750 K, a surface gravity log $g$ between 4.0 and 3.0, and solar 
metallicity, which is confirmed by model fitting to high--resolution spectra 
in selected wavelength ranges (see Fig. 3, Supplementary Fig. 1). For the 
spectral type found (G0--G2) and being a slightly evolved star (surface 
gravity not much below the main--sequence value), the mass should be about 
solar ($M \approx 1$ M$_{\odot}$) and thus the radius, for the range of 
surface gravities above, should be $R \approx 1$--3 R$_{\odot}$,  which 
translates (via our photometric data) into a distance $d \approx 2.5$--4.0
kpc. This companion could have been a main--sequence star or a subgiant before 
the explosion. While main--sequence companions might no longer look like 
ordinary main--sequence stars after the  explosion of the type Ia supernova 
(and they might resemble subgiants, their envelopes having expanded after 
the supernova impact), subgiants would remain subgiants of lower surface 
gravity $^{9,10,14,15}$. 

Stars at distances $d \approx 2$--4 kpc, in that direction, are moving 
at average radial velocity$^{17}$ $v_{r} \approx -20$ to $-40$ km
s$^{-1}$ (in the Local Standard of Rest), with a $\sim$ 20 km s$^{-1}$ 
velocity dispersion$^{18, 19}$. Tycho G moves at $-108 \pm 6$ km s$^{-1}$ 
(heliocentric) in the radial direction. The deviation of Tycho G from the 
average thus exceeds by a factor of 3 the velocity dispersion of its stellar 
type. It has a 0.3\% probability of having that characteristic and being
unrelated to the explosion (that is, it is a 3$\sigma$ outlier).  
In contrast, all other stars with distances compatible with that of 
SN 1572 have radial velocities within the velocity dispersion  
as compared with the average of all stars at the same location in the Galaxy. 
We studied through detailed proper motion measurements on the
{\it HST} WFPC2 images$^{20}$ whether Tycho G has a high tangential velocity 
as well (see Supplementary Table 2 and Supplementary Methods).
Tycho G has significant proper motion toward lower Galactic latitude: 
$\mu_{b} = -6.11 \pm 1.34$ mas yr $^{-1}$ (the proper motion along longitude 
is small, $\mu_{l} = -2.6 \pm 1.34$ mas yr$^{-1}$). 
The proper motion in Galactic latitude implies that this star is 
an outlier in proper motion as well, with a derived tangential velocity of 
$94 \pm 27$ km s$^{-1}$ (a 24 km s$^{-1}$ systematic error was 
added, resulting from a 1.7 mas yr$^{-1}$ uncertainty in the reference 
frame solution of the images). The other stars do not 
show such coincidence in distance and high tangential velocity. 
The modulus of the velocity vector has a value of 136 km s$^{-1}$, 
which is over a factor of 3 larger than the mean velocity value at 3 kpc. 

If Tycho G is the companion star as suggested by its kinematics, 
the explosion centre should have been 2.6 arcsec north of the current location 
of this star on the basis of its velocity. The peculiar velocity would 
correspond to the peculiar velocities expected from the disruption of a white 
dwarf plus subgiant/main--sequence system$^{9,10}$ of roughly a solar mass. 
The system would have resembled the recurrent nova U Scorpii (see 
Supplementary Note 2). The excess velocity corresponds to a period of about 
2--7 days, for a system made of a white dwarf close to the Chandrasekhar 
mass plus a companion of roughly a solar mass at the moment of the explosion. 

Several paths lead to this star as the likely donor star of SN 1572: its high 
peculiar velocity (both radial and tangential velocities), the distance in 
the range of SN 1572, and its type, which fits the post--explosion profile of 
a type Ia supernova companion, as the position of this star in the 
Hertzsprung--Russell diagram is also untypical for a standard subgiant. The 
lower limit to the metallicity obtained from the spectral fits is [M/H] $> 
-0.5$ (see Fig. 3 and Supplementary Fig. 1), which excludes its belonging to 
the Galactic halo population as an alternative explanation of its high 
velocity. Spectra taken at five different epochs also exclude its being a 
single--lined spectroscopic binary. If our candidate is the companion star, 
its overall characteristics imply that the supernova explosion  would affect 
the companion mainly through the kinematics. Our search for the binary 
companion of Tycho's supernova has excluded giant stars. It has also shown 
the absence of blue or highly luminous objects as post--explosion companion 
stars. A star very similar to the Sun but of a slightly more evolved type is 
here suggested as the likely mass donor that triggered the explosion of SN 
1572. That would connect the explosion to the family of cataclysmic variables.

\vfill\eject

\noindent{\bf Supplementary Information} accompanies the paper
on {\bf www.nature.com/nature}. 

\bigskip

\noindent{\bf Acknowledgements}

\noindent
P.R.L. thanks C. Ruiz Ogara for giving her the spirit to complete this 
survey. We thank the support staff at the European Northern Observatory at 
La Palma for their assistance throughout this project, as well as the 
support staff at the W. M. Keck Observatory and NASA/ESA Hubble Space
Telescope. We express our special gratitude to C. Abia, F. Figueras, C. 
Guirao, R. Mignani and J. Torra for diverse consultations. This work has 
been supported by DURSI, DGYCIT (to P.R.L, J.M and R.C), PPARC (to J.M and  
S.J.S) aand by NSF (to A.V.F., R.C., R.J.F. and R.L.K).

\bigskip

\noindent{\bf Correspondence} and requests should be addressed
to P.R.L. (e--mail: pilar@am.ub.es)

\clearpage

\noindent{\bf FIGURE CAPTIONS}

\bigskip

\noindent
{\bf Figure 1} Positions and proper motions of stars. Positions are
compared with three centres: the Chandra (Ch) and 
ROSAT (RO) geometrical centres of the X--ray emission, and that of
the radio emission (Ra). Dashed lines indicate circles of 0.5 arcmin  around  
those centres. The supernov position reconstructed from Tycho Brahe's 
measurements (Ty) is also shown, though merely for its historical 
interest$^{21}$. The radius of the remnant is about 4 arcmin and the SNR is 
quite spherically symmetric, with a fairly good coincidence between radio and 
continuum X--ray emission$^{8,22,23}$. However, there is a 0.56 arcmin 
displacement along the east--west axis between the radio emission and
the high--energy continuum in the 4.5--5.8 keV band observed by 
XMM--Newton in the position of the western rim$^{23}$ (Supplementary 
Note 1). Such asymmetry amounts to a 14\% offset along the east--west axis.  
In SNRs from core--collapse supernovae (type II supernovae), 
up to a 15$\%$ discrepancy between the location of the compact object 
and the geometric center is found in the most symmetric cases$^{24}$.
On the basis of the above considerations, in our search we cover 15\% of
the innermost radius (0.65 arcmin) of the SNR around the Chandra centre 
of SN 1572. The companion star, if there is any, is unlikely to
be outside this area (solid line). 
The proper motions of the stars measured from  HST WFPC2 images are 
represented by arrows, their lengths indicate the total displacements between 
AD 1572 and present. Error bars are shown by parallel segments. Red circles are
the extrapolated positions of the stars back to AD 1572. Star Tycho G 
displays a high proper motion, corresponding to the highest tangential 
velocity in the field, since both stars U and O are at much shorter distances 
(see Supplementary Methods).

\bigskip

\noindent
{\bf Figure 2} The SN 1572 field and radial velocity of the stars. {\bf a}, 
Image from the Auxiliary  Port at the William Herschel Telescope. It confirms 
the relative emptiness of the field. The search area (see also Fig. 1 bold 
circle) covers a radius of 0.65 arcmin around RA = 00 h 25 min 19.9 s, dec. = 
64$^{o}$ 08$^{'}$ 18.2$^{''}$ (J2000) (the Chandra geometrical centre of 
X--ray emission) with repeated photometric and spectroscopic observations of 
the included stars at various epochs to check for variability and exclude 
binarity. Additional stars have been observed outside of the 0.65 arcmin 
radius area and are visible in this field (whose diameter is 1.8 arcmin). For 
a remnant distance $d = 3.0$ kpc and a visual extinction $A_{V} = 1.7$--2.0 
mag toward the candidate stars (see Supplementary Table 3), our search limit 
down to an apparent visual magnitude $V = 22$ implies that the survey must 
have detected all main--sequence stars of spectral types earlier than K6, 
plus all subgiant, giant and supergiant stars within the corresponding cone. 
At that distance and with such extinction, the Sun would shine as a $V = 
18.9$ mag star. {\bf b}, Radial velocity (in the Local Standard of Rest, LSR) 
versus distance for the subsample of stars closer than 6.5 kpc (the other 
stars are at a distance well beyond the SNR). We are looking outward along 
the Galactic plane, and the dashed line shows the approximate relationship 
for the stars in the direction of Tycho given by the expression 
$v_{r} = -v_{\odot}\ {\rm cos}(l - l_{\odot}) + A\ r\ {\rm sin}(2l)$,
where $l$ and $l_{\odot}$ are the respective Galactic longitudes of Tycho 
and the solar apex, $v_{\odot}$ is the Sun's velocity in the LSR, and $A$ is 
Oort's constant$^{18}$. We include two field stars (stars O and U)
that are slightly away from the search area (at $>$15\% of the radius of the 
SNR) but at distances in the range 2--4 kpc as well. (Star names are labelled
lower case in {\bf a} for clarity.)

\bigskip

\noindent
{\bf Figure 3} Model fits to observed spectra. Model atmosphere parameters 
are those listed in Table 1, and chemical abundances are solar. They are 
shown here for our candidate star for the companion of SN 1572 (Tycho G) and 
the red giant (Tycho A) and main--sequence star (Tycho B) nearest to the 
distance of SN 1572 and to the X--ray centre. Identifications of the most 
significant metal lines are given. We have not detected significant 
spectroscopic anomalies, either here or in the whole sample, and most spectra 
are well reproduced assuming solar abundances$^{25}$. Thin lines correspond 
to the observations and thicker lines to the synthetic spectra. Spectra were 
obtained at the William Herschel Telescope (WHT) with UES and ISIS. Tycho A 
(bottom panel) is the closest red giant in the sample. It is a K0 III star, 
and its mass should be typically $M \approx 3$~M$_{\odot}$ (M$_{\odot}$ 
stands for the mass of the Sun). Here, Tycho A is ruled out simply on the 
basis of having too short a distance. All the other red giants are located 
well beyond Tycho's remnant, and  therefore also ruled out (see Supplementary 
Discussion). The A8/A9 star Tycho B (second panel from bottom) has $M 
\approx 1.5$~M$_{\odot}$, which would fall within the appropriate range for 
main--sequence type Ia supernova companions, as it would have been massive 
enough to transfer the required amount of mass to the WD. The entirely normal 
atmospheric parameters, however, strongly argue against any such event in the 
star's recent past. The low radial velocity reinforces this conclusion. 
Tycho G (three upper panels). The second and third spectra from the top show
computed spectra compared with observed spectra obtained at the WHT with 
ISIS. The upper panel shows the observed spectrum near H${\alpha}$. This  
line is blueshifted, implying a peculiar radial velocity exceeding about 3 
times the velocity dispersion for its stellar type. This star does not 
belong to the halo population (Supplementary Fig. 1).

\vfill\eject

\begin{table*}[htb]
\label{table:1}
\newcommand{\m}{\hphantom{$-$}}
\newcommand{\cc}[1]{\multicolumn{1}{c}{#1}}
\caption{\bf Table 1. Characteristics of the supernova companion candidates}
\begin{tabular}{@{}lllllll}
\\
\hline

Star    & $\theta$ & Spec. type & T$_{eff}$ & log $g$ & E(B--V) & d \\
        & (arcsec) & \& lum. class &(K)  & (c.g.s.)& (mag.) & (kpc) \\
\hline
\\
Tycho A & 1.6 & K0--K1 III  &  4750  & 2.5$^{+0.5}_{-0.5}$ &   
 0.55$^{+0.05}_{-0.05}$& 1.1$^{+0.3}_{-0.3}$ \\
\\
Tycho B & 1.5 & A8--A9 V &  7500  & 4.5$^{+0.5}_{-0.5}$ & 
 0.60$^{+0.05}_{-0.05}$ & 2.6$^{+0.5}_{-0.5}$ \\ 
\\
Tycho C1 & 6.5 & K7 V &  4000 & 4.5$^{+0.5}_{-0.5}$ & 
 0.5$^{+0.1}_{-0.1}$ & 0.75$^{+0.5}_{-0.5}$ \\

Tycho C2 & 6.5 & F9 III &  6000 & 2.0$^{+0.5}_{-0.5}$ & 
  0.6$^{+0.1}_{-0.1}$ & $>$ 20 \\
\\
Tycho D & 8.4  & M1 V & 3750 & 4.5$^{+0.5}_{-0.5}$ &  
  0.6$^{+0.3}_{-0.3}$ & 0.8$^{+0.3}_{-0.2}$ \\
\\
Tycho E & 10.6 & K2--K3 III & 4250 & 2.0$^{+0.5}_{-0.5}$ &  
 0.60$^{+0.10}_{-0.10}$ & $>$ 20 \\
\\
Tycho F & 22.2 & F9 III & 6000 & 2.0$^{+0.5}_{-0.5}$ &  
 0.54$^{+0.22}_{-0.22}$ & $>$ 10 \\
\\
Tycho G & 29.7 & G2 IV  & 5750 & 3.5$^{+0.5}_{-0.5}$ &  
 0.60$^{+0.05}_{-0.05}$ & 3.0$^{+1.0}_{-0.5}$ \\
\\
Tycho H & 30.0 & G7 III & 5000 & 3.0$^{+0.5}_{-0.5}$ &  
 0.60$^{+0.09}_{-0.09}$ & $>$ 13 \\
\\
Tycho J & 33.9 & K1 V & 5000 & 4.5$^{+0.5}_{-0.5}$ &  
 0.58$^{+0.12}_{-0.11}$ & 2.4$^{+0.3}_{-0.2}$ \\
\\
Tycho K & 35.0 & F9 III & 6000 & 2.0$^{+0.5}_{-0.5}$ &
 0.60$^{+0.10}_{-0.10}$ & $>$ 10 \\
\\
Tycho N & 35.4 & G0 V & 6000 & 4.5$^{+0.5}_{-0.5}$ &  
 0.62$^{+0.08}_{-0.07}$ & 2.1$^{+0.7}_{-0.7}$ \\
\\
Tycho V & 29.2 & K3 V & 4750 & 4.5$^{+0.5}_{-0.5}$ & 
 0.60$^{+0.10}_{-0.10}$ & 3.8$^{+0.6}_{-0.6}$ \\
\hline
\end{tabular}\\[2pt]
\end{table*}

\noindent{\bf TABLE CAPTION}

\bigskip
\noindent
Supernova companion candidates within the search radius and limiting 
magnitude. Angular distances $\theta$ are from the Chandra X--ray geometrical 
centre, located at RA = 00 h 25 min 19.9 s,  dec.  = 64$^{o}$ 08$^{'}$ 
18.2$^{''}$  (J2000). Synthetic spectra, under the assumption of local 
thermodynamic equilibrium (LTE), are fitted to the observed ones using 
the grids of model atmospheres and the atomic data of Kurucz$^{25}$, with 
the Uppsala Synthetic Spectrum Package$^{27}$. This determines the 
atmospheric parameters effective temperature $T_{eff}$ and surface gravity 
$g$. Intrinsic colours and absolute visual magnitudes are deduced from the 
relationships between spectral type and colour and between spectral type and 
absolute magnitude for the different luminosity classes$^{28}$. Comparison 
with our photometric $BVR$ measurements (see Supplementary Table 3) yields the 
reddening $E(B-V)$, from which the visual extinction $A_{V}$ and the 
corrected apparent visual magnitude ${V_{0}}$ are calculated. Comparison with 
the absolute visual magnitude then gives the distance $d$. Uncertainties in 
$T_{eff}$ are 250 K. Tycho J is a binary of main--sequence stars with masses 
in the range 0.80--0.85 M$_{\odot}$ and quite similar atmospheric parameters. 
Tycho C is found in HST images as being two stars (C1 and C2)  0.25 arcsec
apart. Modelling of the composite spectrum and the HST magnitudes of the 
stars show that they do not constitute a physical binary; the hot
fainter component C2 is at larger distance than C1. Within this list of 
stars, D, G, N and V have proper motions along Galactic longitude and 
latitude of $\mu_{l}$ = --3.23, $\mu_{b}$ = --0.58 $\pm$ 0.66 (same error 
in both coordinates, units in mas yr$^{-1}$) for star D,
$\mu_{l}$ = --2.60, $\mu_{b}$ = --6.11 $\pm$ 1.34 for star G,
$\mu_{l}$ =   3.23, $\mu_{b}$ =   1.45 $\pm$ 1.15 for star N,
and $\mu_{l}$ =   1.61, $\mu_{b}$ = --2.85 $\pm$ 0.78 for star V.

\bigskip

\clearpage
\begin{figure}[hbtp]
\input epsf
\centerline{\epsfysize16cm\epsfbox{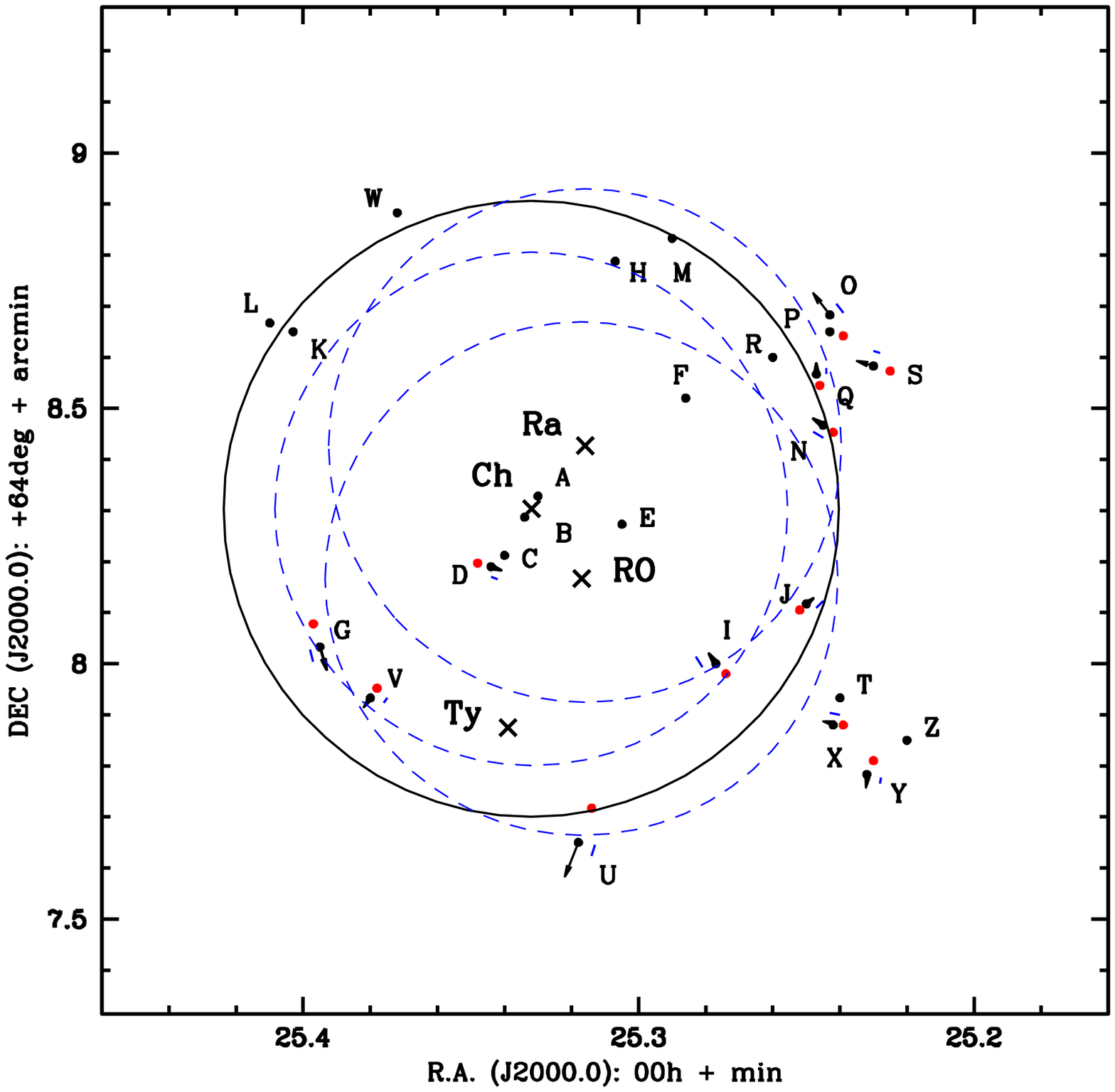}}
\nopagebreak[4]
\caption{}
\end{figure}

\clearpage
\begin{figure}[hbtp]
\input epsf
\hskip104pt{\epsfysize14.0cm\epsfbox{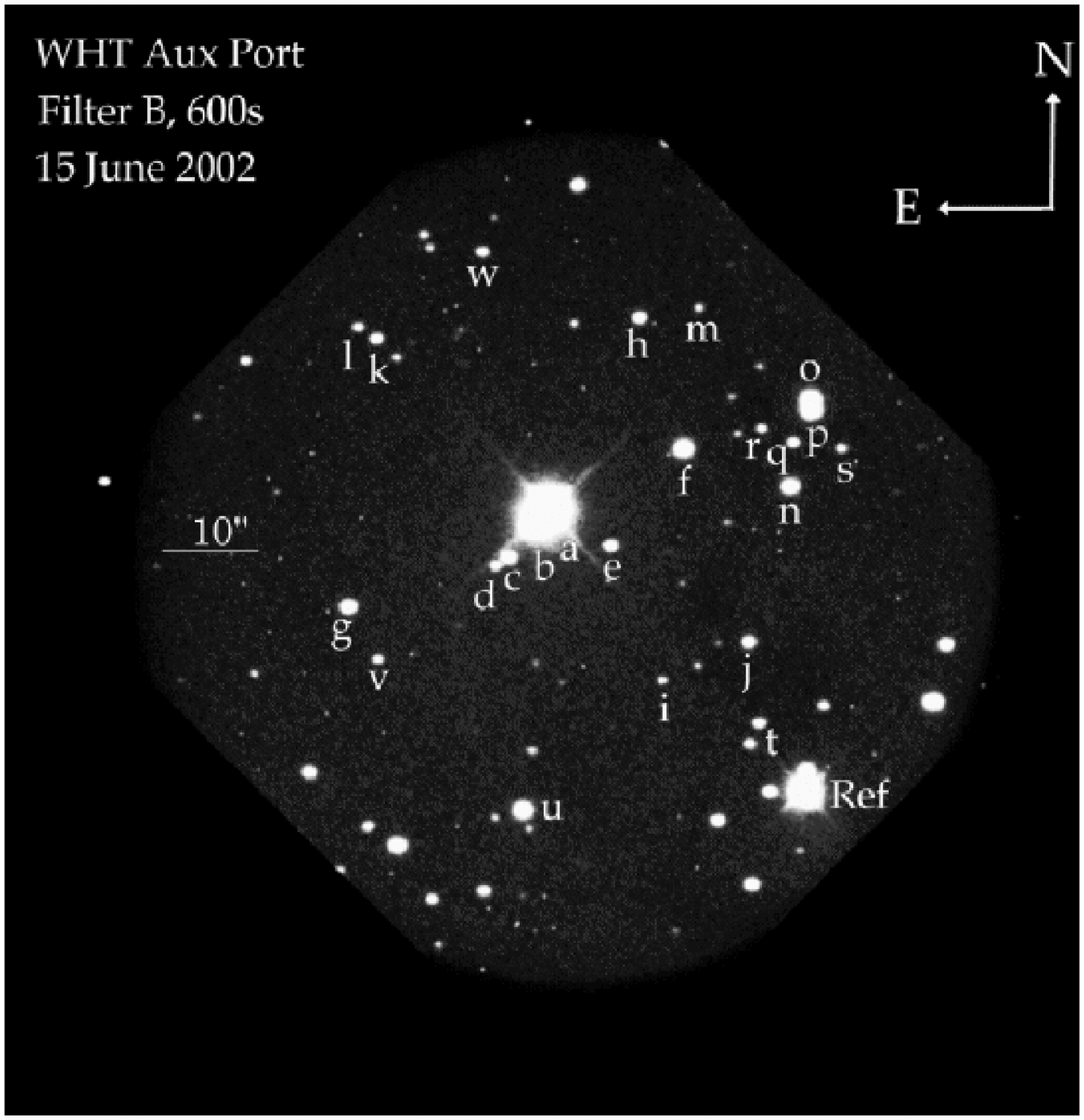}}
\vskip -55pt
\centerline{\epsfysize11.75cm\epsfbox{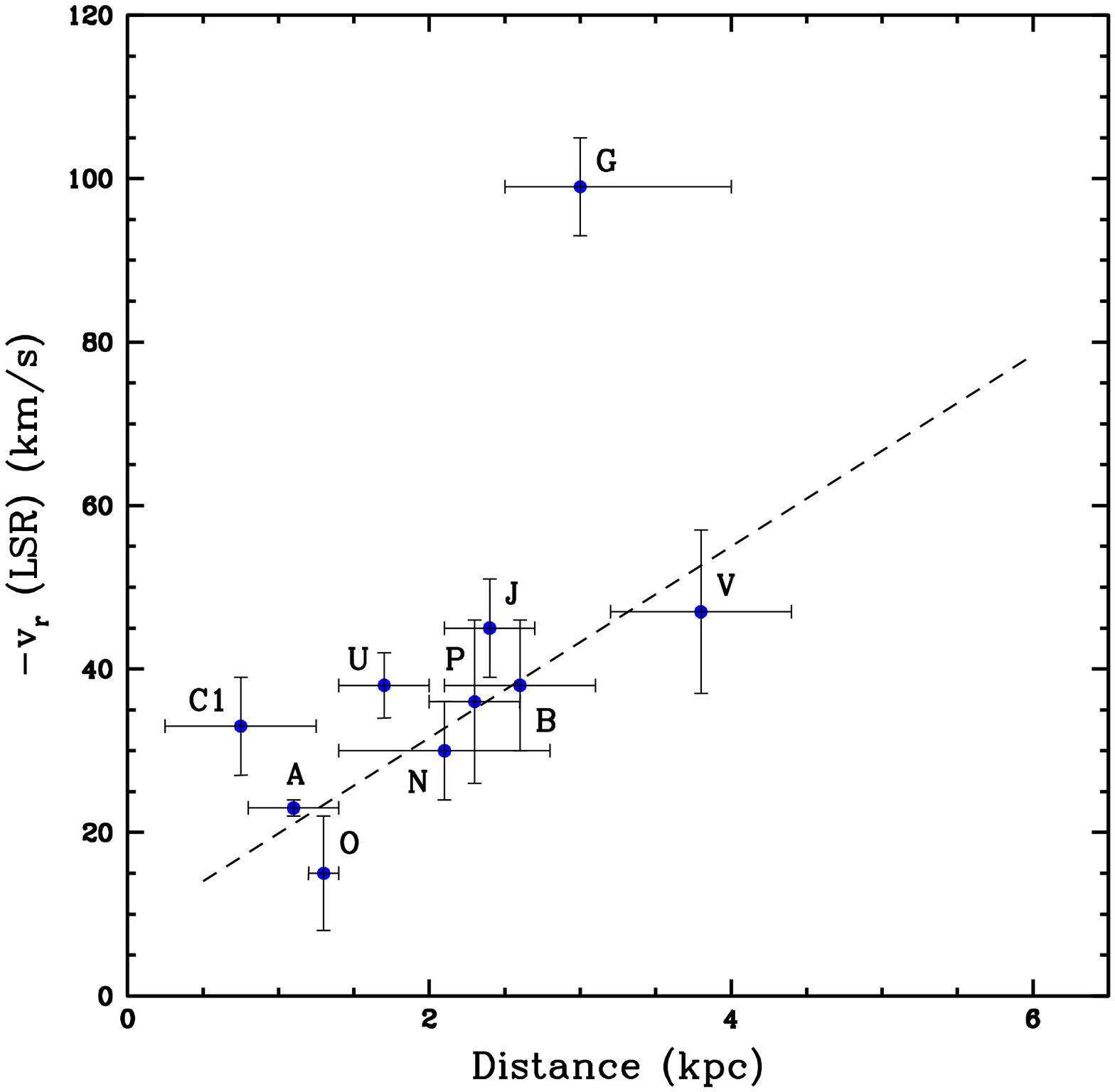}}
\nopagebreak[4]
\label{fig2}
\caption{}
\end{figure}

\clearpage

\begin{figure}[hbtp]
\input epsf
\centerline{\epsfysize16cm\epsfbox{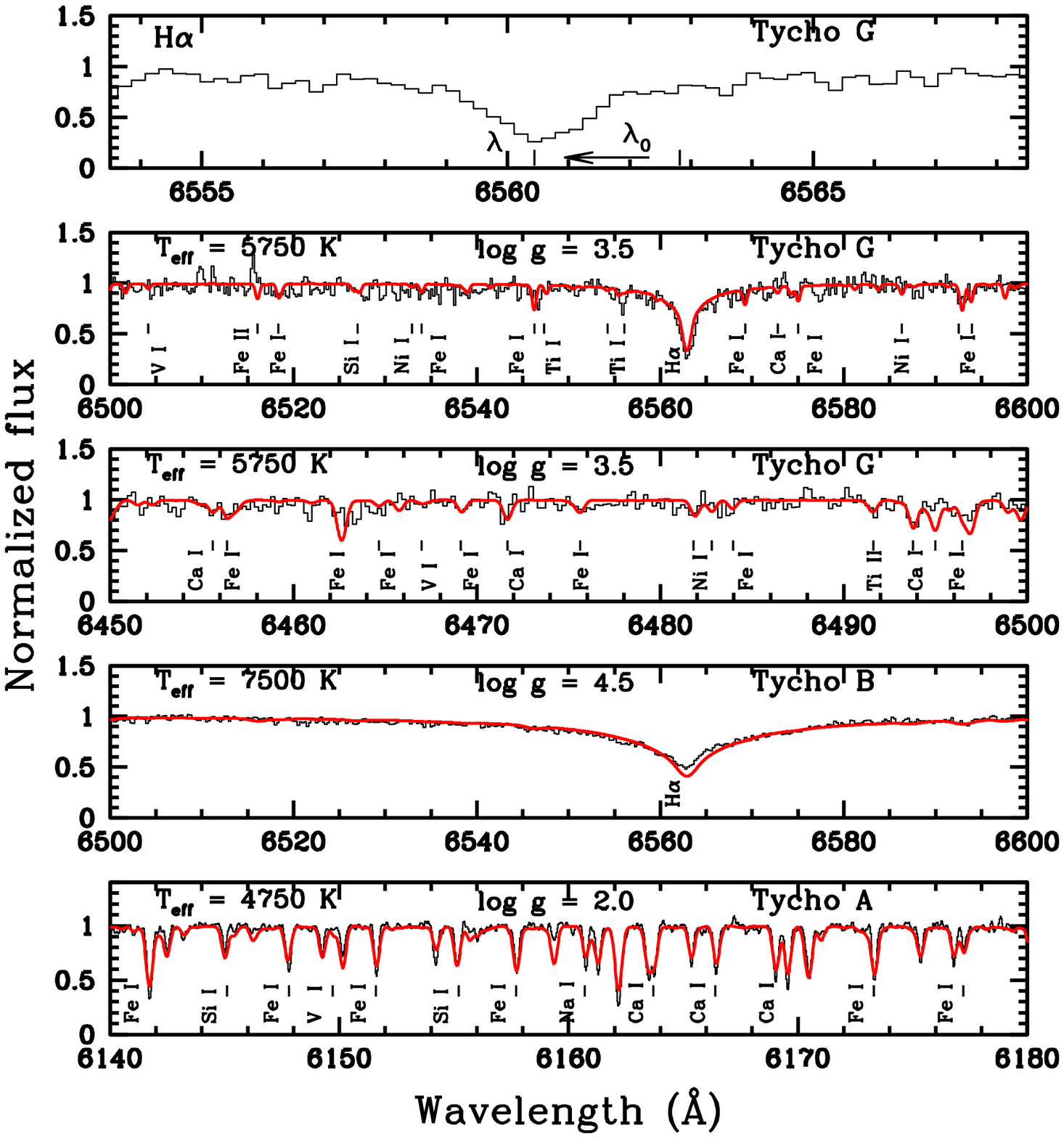}}
\nopagebreak[4]
\caption{}
\label{fig3}
\end{figure}

\clearpage

\vfill\eject

\noindent
{\bf Supplementary Information} 

\smallskip

\noindent
{\bf Supplementary Methods}
Specification of the WFPC2 measurements

\smallskip

\noindent
{\bf Supplementary Table 1}
Spectroscopic runs 
at the 4.2~m William Herschel Telescope (WHT), the 2.5~m Nordic Optical
Telescope (NOT) and the 10~m Keck I and Keck II telescopes used in this work

\smallskip

\noindent
{\bf Supplementary Table 2} 
Hubble Space Telescope data sets used for proper motions measurements

\smallskip

\noindent
{\bf Supplementary Table 3}
Magnitudes and radial velocities of the stars 

\smallskip 

\noindent
{\bf Supplementary Note 1} Note on the origin of the asymmetry in
Tycho SNR

\smallskip

\noindent
{\bf Supplementary Discussion}  Further discussion on red giant and
main sequence stars

\noindent
{\bf Supplementary Figure 1} Spectra of Tycho G showing 
it to be a subgiant not belonging to the halo population

\smallskip

\noindent
{\bf Supplementary Note 2} Tycho Brahe SN as a U Sco system

\vfill\eject

\begin{table}
\begin{center}
{{\bf Supplementary Table 1.} Spectroscopic runs in this work}
\bigskip
\renewcommand{\tabcolsep}{0.8pc} 
\begin{tabular}{lllll}
\hline
\hline
Date &  Telescope & Instrument/ Grating  &  Resolution  & Coverage (\AA)  \\
\hline
1997-12-04  &  WHT  & ISIS+R158B/R158R   &     600      & 3200--9000 \\
1998-01-03  &  WHT  & ISIS+R300B/R158R   &  1700--600   & 3200--9000 \\
1999-02-24  &  WHT  & ISIS+R300B/R158R   &  1700--600   &  3200--9000 \\
2000-11-21  &  Keck I & ESI                &  10000--7000 &  4000--10000  \\   
2001-06-16  &  WHT &  UES                &  50000       & 4000--7100  \\
2001-06-17  &  WHT & ISIS+H2400B/R1200R   & 10000--7000  & 4000--7000  \\
2002-06-12  &  WHT & ISIS+H2400B/R1200R   & 10000--7000  & 4400--7000  \\
2002-06-13  &  WHT & ISIS+H2400B/R1200R   & 10000--7000  & 4400--7000  \\
2002-06-14  &  WHT & ISIS+R300B/R158R     & 1700--600   & 3200--9000  \\
2002-10-09  &  NOT & ALFOSC+Grism9        & 4500--3900   & 4400--9000  \\
2002-12-05  &  WHT & ISIS+H2400B/R1200R   & 10000--7000 & 4800--7000  \\
2002-12-06  &  WHT & ISIS+H2400B/R1200R   & 10000--7000 & 4400--7000  \\
2003-07-27  &  Keck II & LRIS+300/5000    & 850        & 3200--7900  \\
2003-07-30  &  WHT & ISIS+R1200B/R1200R   & 7000 & 4000--7000  \\
2003-07-31  &  WHT & ISIS+R1200B/R1200R   & 7000 & 4000--7000  \\
2003-11-16  &  WHT & ISIS+R1200B/R1200R   & 7000 & 4000--7000  \\
2003-11-17  &  WHT & ISIS+R1200B/R1200R   & 7000 & 4000--7000  \\
2003-11-18  &  WHT & ISIS+R1200B/R1200R   & 7000 & 4000--7000  \\
2003-12-20  &  Keck I & LRIS+150/7500    & 500         & 3156--9400  \\

\hline
\end{tabular}
\end{center}
\noindent
Spectroscopic runs at the 4.2~m William Herschel Telescope (WHT), 
the 2.5~m Nordic Optical Telescope (NOT) and the 10~m Keck I and Keck II 
telescopes used in this work
\end{table}

\vfill\eject

\begin{table}
\begin{center}
{{\bf Supplementary Table 2.} Hubble Space Telescope data sets used for
proper motion measurements}
\bigskip
\renewcommand{\tabcolsep}{0.85pc} 
\renewcommand{\arraystretch}{0.75} 
\begin{tabular}{lllcc}
\\
\hline
\hline
Dataset   &  Date       & Filter  & Exposure Time  & Instrument
\\ 
\hline
U8S9010HM & 2003-11-08  &  F555W  & 100.0 s  & WFPC2 \\
U8S9010IM & 2003-11-08  &  F555W  & 100.0 s  & WFPC2 \\
U8S9010JM & 2003-11-08  &  F555W  & 100.0 s  & WFPC2 \\
U8S9010KM & 2003-11-08  &  F555W  & 100.0 s  & WFPC2 \\
U8S9010LM & 2003-11-08  &  F555W  & 100.0 s  & WFPC2 \\
U8S90201M & 2003-11-06  &  F555W  & 400.0 s  & WFPC2 \\
U8S90202M & 2003-11-06  &  F555W  & 400.0 s  & WFPC2 \\
U8S90203M & 2003-11-06  &  F555W  & 400.0 s  & WFPC2 \\
U8S90204M & 2003-11-06  &  F555W  & 400.0 s  & WFPC2 \\
U8S90205M & 2003-11-06  &  F555W  & 1.0 s    & WFPC2 \\
U8S90206M & 2003-11-06  &  F555W  & 1.0 s    & WFPC2 \\
U8S90207M & 2003-11-06  &  F555W  & 1.0 s    & WFPC2 \\
U8S90208M & 2003-11-06  &  F555W  & 1.0 s    & WFPC2 \\
U8S90209M & 2003-11-06  &  F555W  & 1.0 s    & WFPC2 \\
U8S9020AM & 2003-11-06  &  F555W  & 1.0 s    & WFPC2 \\
U8S9020BM & 2003-11-06  &  F555W  & 0.2 s    & WFPC2 \\
U8S9020CM & 2003-11-06  &  F555W  & 0.2 s    & WFPC2 \\
U8S9020DM & 2003-11-06  &  F555W  & 0.2 s    & WFPC2 \\
U8S9020EM & 2003-11-06  &  F555W  & 0.2 s    & WFPC2 \\
U8S9020FM & 2003-11-06  &  F555W  & 0.2 s    & WFPC2 \\

\hline
\end{tabular}
\end{center}
\end{table}

\begin{table}
\begin{center}
\renewcommand{\tabcolsep}{0.85pc} 
\renewcommand{\arraystretch}{0.75} 
\begin{tabular}{lllcc}
\\
\hline
\hline
Dataset   &  Date       & Filter  & Exposure Time  & Instrument
\\ 
\hline
U8S9010HM & 2003-11-08  &  F555W  & 100.0 s  & WFPC2 \\
U8S9010IM & 2003-11-08  &  F555W  & 100.0 s  & WFPC2 \\
U8S9010JM & 2003-11-08  &  F555W  & 100.0 s  & WFPC2 \\
U8S9010KM & 2003-11-08  &  F555W  & 100.0 s  & WFPC2 \\
U8S9010LM & 2003-11-08  &  F555W  & 100.0 s  & WFPC2 \\
U8S90201M & 2003-11-06  &  F555W  & 400.0 s  & WFPC2 \\
U8S90202M & 2003-11-06  &  F555W  & 400.0 s  & WFPC2 \\
U8S90203M & 2003-11-06  &  F555W  & 400.0 s  & WFPC2 \\
U8S90204M & 2003-11-06  &  F555W  & 400.0 s  & WFPC2 \\
U8S90205M & 2003-11-06  &  F555W  & 1.0 s    & WFPC2 \\
U8S90206M & 2003-11-06  &  F555W  & 1.0 s    & WFPC2 \\
U8S90207M & 2003-11-06  &  F555W  & 1.0 s    & WFPC2 \\
U8S90208M & 2003-11-06  &  F555W  & 1.0 s    & WFPC2 \\
U8S90209M & 2003-11-06  &  F555W  & 1.0 s    & WFPC2 \\
U8S9020AM & 2003-11-06  &  F555W  & 1.0 s    & WFPC2 \\
U8S9020BM & 2003-11-06  &  F555W  & 0.2 s    & WFPC2 \\
U8S9020CM & 2003-11-06  &  F555W  & 0.2 s    & WFPC2 \\
U8S9020DM & 2003-11-06  &  F555W  & 0.2 s    & WFPC2 \\
U8S9020EM & 2003-11-06  &  F555W  & 0.2 s    & WFPC2 \\
U8S9020FM & 2003-11-06  &  F555W  & 0.2 s    & WFPC2 \\
U8S9020GM & 2003-11-06  &  F555W  & 0.2 s    & WFPC2 \\
U8S9020HM & 2003-11-06  &  F555W  & 100.0 s  & WFPC2 \\
U8S9020IM & 2003-11-06  &  F555W  & 100.0 s  & WFPC2 \\
U8S9020JM & 2003-11-06  &  F555W  & 100.0 s  & WFPC2 \\
U8S9020KM & 2003-11-06  &  F555W  & 100.0 s  & WFPC2 \\
U8S9020LM & 2003-11-06  &  F555W  & 100.0 s  & WFPC2 \\

\hline
\end{tabular}
\end{center}
\noindent
{\bf Supplementary Table 2 (cont.)} The data from 1999
correspond to {\it HST} programmes {\it GO}6435 and {\it GO}7405. The
data from 2003 correspond to {\it GO}9729.
\end{table}

\vfill\eject

\begin{table}
\renewcommand{\tabcolsep}{1.5pc} 
\begin{center}
{\bf Supplementary Table 3. Magnitudes and radial velocities of the
stars}

\bigskip

\begin{tabular}{cccccc}\\
\hline
\hline
Star & $\theta$ & B       & V       & R       & v$_{r}$ \\
     & (arcsec) & (mag.)  & (mag.)  & (mag.)  & (km/s)  \\
\hline
A    & 1.6      & 14.82$^{+0.03}_{-0.03}$ & 13.29$^{+0.03}_{-0.03}$          
& 12.24$^{+0.03}_{-0.03}$ & --23$^{+1}_{-1}$  \\ 
B    & 1.5      & 16.35$^{+0.03}_{-0.03}$ & 15.41$^{+0.03}_{-0.03}$ 
& 15.11$^{+0.10}_{-0.10}$ & --38$^{+8}_{-8}$  \\
C$^{*}$   & 6.5      & 21.06$^{+0.12}_{-0.12}$ & 19.06$^{+0.05}_{-0.05}$ 
& 17.77$^{+0.03}_{-0.03}$ & --33$^{+6}_{-6}$  \\
D    & 8.4      & 22.97$^{+0.28}_{-0.28}$ & 20.70$^{+0.10}_{-0.10}$          
& 19.38$^{+0.06}_{-0.06}$ & ---               \\
E    & 10.6     & 21.24$^{+0.13}_{-0.13}$ & 19.79$^{+0.07}_{-0.07}$          
& 18.84$^{+0.05}_{-0.05}$ & --26$^{+18}_{-18}$ \\
F    & 22.2     & 19.02$^{+0.05}_{-0.05}$ & 17.73$^{+0.03}_{-0.03}$          
& 16.94$^{+0.03}_{-0.03}$ & --34$^{+11}_{-11}$ \\
G    & 29.7     & 20.09$^{+0.08}_{-0.08}$ & 18.71$^{+0.04}_{-0.04}$          
& 17.83$^{+0.03}_{-0.03}$ & --99$^{+6}_{-6}$   \\ 
H    & 30.0     & 21.39$^{+0.14}_{-0.14}$ & 19.80$^{+0.07}_{-0.07}$          
& 18.78$^{+0.05}_{-0.05}$ & --71$^{+10}_{-10}$ \\
J    & 33.9     & 21.15$^{+0.12}_{-0.12}$ & 19.74$^{+0.07}_{-0.07}$          
& 18.84$^{+0.05}_{-0.05}$ & --45$^{+6}_{-6}$   \\
K    & 35.0     & 21.64$^{+0.15}_{-0.15}$ & 20.11$^{+0.08}_{-0.08}$          
& 19.15$^{+0.05}_{-0.05}$ & --33$^{+10}_{-10}$ \\
N    & 35.4     & 19.59$^{+0.06}_{-0.06}$ & 18.29$^{+0.04}_{-0.04}$          
& 17.47$^{+0.03}_{-0.03}$ & --30$^{+6}_{-6}$   \\
V    & 29.2     & 23.32$^{+0.33}_{-0.33}$ & 21.41$^{+0.13}_{-0.13}$          
& 20.20$^{+0.08}_{-0.08}$ & --47$^{+10}_{-10}$ \\
\hline
O    & 41.5     & 18.62$^{+0.04}_{-0.04}$ & 17.23$^{+0.03}_{-0.03}$          
& 16.37$^{+0.03}_{-0.03}$ & --15$^{+7}_{-7}$   \\
P    & 40.4     & 18.84$^{+0.10}_{-0.10}$ & 17.61$^{+0.03}_{-0.03}$          
& 16.78$^{+0.03}_{-0.03}$ & --36$^{+10}_{-10}$ \\
U    & 39.5     & 19.03$^{+0.05}_{-0.05}$ & 17.73$^{+0.03}_{-0.03}$          
& 16.95$^{+0.03}_{-0.03}$ & --38$^{+4}_{-4}$   \\
\hline
\hline
\end{tabular}\\[2pt]
\end{center}
\noindent
Angular distances from Chandra's geometrical X--ray centre, BVR 
apparent magnitudes and radial velocities (LSR) for the sample of SN 
companion candidates. Radial velocities have been measured from the
wavelength shifts of several absorption lines in each observed
spectrum (see Supplementary Table 1).
$^{*}$Data are for the unresolved pair. From the HST data, the brighter, bluer
component has magnitudes B = 21.28, V = 19.38, R = 18.10 while the
fainter, redder component has B = 22.91, V = 20.53, R = 19.23. 
\end{table}

\vfill\eject

\noindent
{\bf Supplementary Methods}

\smallskip

\noindent
{\bf Specification of the WFPC2 measurements}

\smallskip

The proper motions were 
calculated by comparing through a maximum likelihood analysis  
the displacement in image centroids of the 
target stars to a reference frame determined from all the point sources 
in common between the WFPC2 images of the two epochs$^{1}$.

\smallskip

\noindent
For the majority of our targets, the stellar images fell on the WF3 CCD 
of the WFPC2 mosaic camera on both epochs. However, due to
differences in the camera pointing and roll-angle between the exposures 
at the two epochs, the target star K fell on different 
CCDs of the WFPC2 mosaic at the two epochs, as did the
stars L, M, and W (these three not being targets). The overlap region proved 
to be too small to determine a reference frame, and so the proper 
motions of these stars could not be measured. A similar
effect entered the measurement of target H, which had to be done using images
spanning a shorter time baseline, and making the uncertainty too 
large. No accurate
image centroids were derived for the stars C1 and C2 due to their small
separation, neither for
 Target A which is saturated in the reference images and has a  
background star at half a second of its centroid.
The other stars do not have 
significant proper motions (at $>$ 2$\sigma$). This proper
motion programme continues in {\it HST} Cycle 13 where measurements with
smaller error bars will be
obtained using both WFPC2 and ACS.

\smallskip

\noindent
$^{1}$ Ibata, R. A. \& Lewis, G. F.,
Proper motion measurements with WFPC. 
 {Astron. J}, {\bf 116}, 2569 (199

\vfill\eject

\noindent
{\bf Supplementary Note 1}

\smallskip

\noindent
{\bf Note on the origin of the asymmetry in Tycho SNR}

Evidence 
that the ejecta encountered a
dense H cloud at the eastern edge
giving rise to brighter emission and lower ejecta velocity there, while 
finding a lower-density 
medium in the western rim, might account for the
asymmetry$^{1}$.

To obtain an approximate position of the dynamical center of the explosion
one would need to trace back the expansion 
of the SNR to the time of the explosion from the SNR expanding
filaments. Unlike in SNe~II, 
neither in SN 1572 nor in any other SN~Ia has the 
derivation been possible thus far, due to the faintness of 
the filaments. This measurement remains an interesting long--term project to
be undertaken with the use of large telescopes. In this work, 
arguments based on what is found in  SNe~II
together with the E--W geometrical asymmetry of this SNIa are used to
establish the area of the search. 

\noindent
$^{1}$ Decourchelle, A. et al. 
XMM-Newton observation of the Tycho supernova remnant
{\it Astron \& Astrophys. } {\bf 365}, L218--L224 (2001).

\vfill\eject

\noindent
{\bf Supplementary Discussion}

\smallskip

\noindent
{\bf Further discussion on red giants and main sequence stars}

\noindent
{\it Red giants} In the red-giant companion case, the envelope is 
loosely bound gravitationally, and 
upon collision with the SN~Ia ejecta it should be 
either completely stripped or 
just a small fraction of it remain bound to the core$^{1,2}$. If the full 
envelope were lost, the remaining He core would appear as a hot He pre-white 
dwarf, not as a red giant. 
Our survey would have revealed the 
presence of hot helium stars resulting from complete 
stripping of the envelope of a red giant because they are brighter
than the other stars that we do observe. 
If a minor fraction of the envelope of the red 
giant were retained instead, the H-burning shell would remain active
and the residual envelope would expand to red-giant size.
In the case of Tycho A, whose envelope is $\sim 
2.6$~M$_{\odot}$, nothing similar to the star that we are now seeing could 
ever be what remains after the supernova impact. 
Tycho A and the other red giants are in any case 
ruled out on the basis of their incompatible distance with SN 1572. 

\noindent
{\it Main sequence stars} 
We should emphasize that there exists, beyond differences in the 
post-explosion evolution calculations, unanimous agreement that the companion 
object, if it were a main-sequence star, should exhibit an odd combination 
of stellar parameters (log~$g$ and $T_{eff}$) for its mass and that it should 
move fast$^{1,3,4,5}$. That is not found in the observed sample of 
main-sequence stars. Apart from Tycho B (see Figure 3), there are six 
main-sequence stars in Table 1. One of them (Tycho J) is a spectroscopic 
binary made of two main-sequence stars with similar spectral types 
(G8/K0-K3). Its radial velocity fits the derived distance, and no 
such system is likely to be the surviving companion of any SN Ia. As stated 
above, all main-sequence stars of spectral type earlier than K6 were
detected to the distance of the remnant. The total mass available for 
transfer in main-sequence stars of type 
later than K6 ($M \lapprox 0.6$~M$_{\odot}$) excludes these stars as 
viable candidates to SN Ia companions.

\noindent
$^{1}$ Canal, R., M\'endez, J. \& Ruiz-Lapuente, P. Identification of the 
companion stars of Type Ia supernovae. {\it Astrophys. J.} {\bf 550}, 
L53--L56 (2001)

\noindent
$^{2}$ Livne, E., Tuchman, Y. \& Wheeler, J.C. Explosion of a supernova with a 
red giant companion. {\it Astrophys. J.} {\bf 399}, 665--671 (1992)

\noindent
$^{3}$ Ruiz-Lapuente, P., Comeron, F., Smartt, S., Kurucz, R., 
M\'endez, J., Canal, R., Filippenko, A., Chornock, R. 
Search for the companions of Galactic SNe~Ia, in {\it 
From Twilight to Highlight: the Physics of Supernovae}, 140, ed.
W. Hillebrandt \& B. Leibundgut
(Berlin: Springer-Verlag) (2003).

\noindent
$^{4}$ Marietta, E., Burrows, A. \& Fryxell, B. Type Ia supernova
 explosions in 
binary systems: the impact on the secondary star and its consequences. 
{\it Astrophys. J. Suppl.} {\bf 128}, 615--650 (2000). 

\noindent
$^{5}$ Podsiadlowski, P. On the evolution and appearance of a surviving 
companion after a Type Ia supernova explosion,
astro-ph/0303660  (2003).

\vfill\eject

\begin{figure}[hbtp]
\input epsf
\centerline{\epsfysize16cm\epsfbox{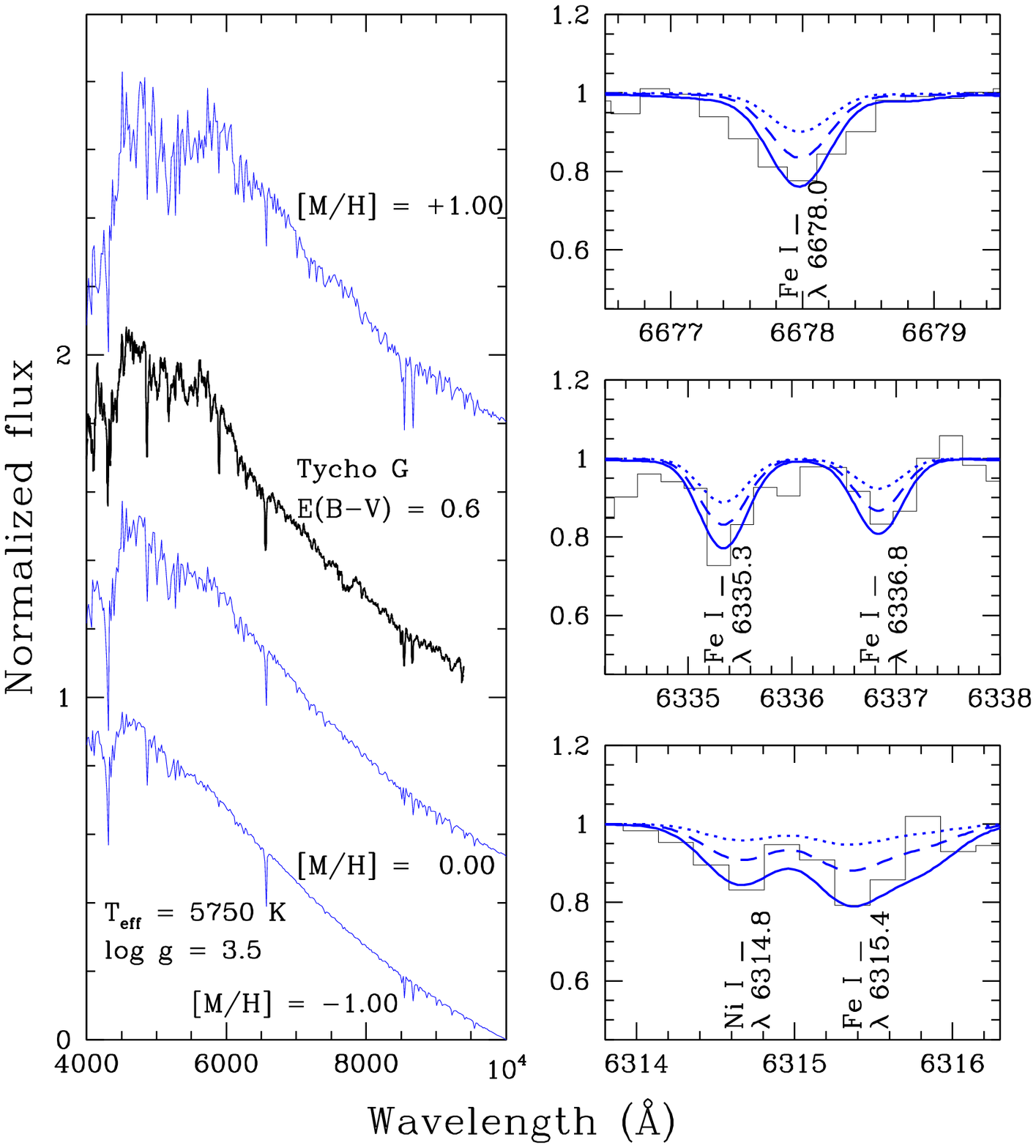}}
\nopagebreak[4]
\end{figure}

\smallskip

\noindent
{\bf Supplementary Figure 1}. Spectra of Tycho G showing 
it to be a subgiant not belonging to the halo population. (Right panel) 
Several fits to Fe and Ni lines in Tycho G for solar abundances (bold) and
abundances [Fe/H] $= -0.5$ (dashed line) and  [Fe/H] $=  -1$ (dotted line). 
This star does not belong to the halo population: it shows solar metallicities 
in Fe and Ni. The observed spectra were obtained with
ISIS at the WHT.  (Left panel) A low-resolution spectrum
 over a wide wavelength 
range was obtained with LRIS at the Keck Observatory (second from top) 
and it is compared with template model spectra of the same spectral class
and various metallicities. It
has also been used to determine the atmospheric parameters (see 
main text). Very high and low metallicities are excluded. The
spectrum of Tycho G confirms that it is a subgiant star.

\vfill\eject

\noindent
{\bf Supplementary Note 2}

\smallskip

\noindent
{\bf Tycho Brahe SN as a U Sco system}

\smallskip

\noindent
The evolutionary path that gave rise to SN 1572  could be 
similar to that leading to the recurrent nova U Scorpii pointed out as
a candidate to SN Ia progenitor. The companion candidate is now a mildy 
evolved star within the solar mass range. The excess velocity implies
an orbit of 6 days before the explosion. Starting from a white dwarf
with a mass $\sim$ 0.8 M$_{\sun}$ plus a somewhat evolved companion of 
$\sim$ 2.0-2.5 M$_{\sun}$ filling its Roche lobe, and with a period 
of 12 days, it would have ended up as a white dwarf at the Chandrasekhar 
mass ($\sim$ 1.4 M$_{\sun}$) plus a companion of roughly 1 M$_{\sun}$, the 
period then being $\sim$ 6 days (orbital velocity $\sim$ 90 km/s). The 
effective radius of the Roche lobe of the companion just before explosion 
would have been $\sim$ 7 R$_{\sun}$. Now the radius of the companion has 
to be less than three times the solar radius given the effective temperature 
and luminosity of the star. The smaller current radius would result from 
the opposite effects of mass stripping and shock heating by the supernova 
impact, plus subsequent fast cooling of the outer layers up to the present 
time.

\end{document}